\documentclass[a4,12pt]{article}
\usepackage{cite,type1cm,amsmath,amssymb,color,graphics,amscd,amsfonts,epsf,indentfirst}
\usepackage{bm,type1cm,amsmath,amssymb,color,amscd,amsfonts,epsf,indentfirst}
\usepackage{epsfig}
\usepackage{subfigure}
\usepackage{graphicx} 
\usepackage{enumerate}
\usepackage[dvipdfm,hypertex]{hyperref}
\allowdisplaybreaks[4]

\makeatletter
\@addtoreset{equation}{section}

\makeatletter
\renewcommand\section{\@startsection {section}{1}{\z@}%
                                   {-3.5ex \@plus -1ex \@minus -.2ex}
                                   {2.3ex \@plus.2ex}%
                                   {\normalfont\large\bfseries}}
\renewcommand\subsection{\@startsection{subsection}{2}{\z@}%
                                     {-3.25ex\@plus -1ex \@minus -.2ex}%
                                     {1.5ex \@plus .2ex}%
                                    {\normalfont\bfseries}}

\begin{document}
\begin{titlepage}
  \thispagestyle{empty}
  
  \begin{flushright} 
    YITP-10-35
  \end{flushright} 
  
  \vspace{2cm}
  
  \begin{center}
    \font\titlerm=cmr10 scaled\magstep4
    \font\titlei=cmmi10 scaled\magstep4
    \font\titleis=cmmi7 scaled\magstep4
     \centerline{\titlerm
Non-equilibrium Condensation Process}
    \vspace{0.8cm}
\centerline{\titlerm
in a Holographic Superconductor         }
    
    \vspace{2.2cm}
    \noindent{{
        Keiju Murata\footnote[1]{K.Murata@damtp.cam.ac.uk}, 
        Shunichiro Kinoshita\footnote[2]{kinosita@yukawa.kyoto-u.ac.jp} 
        and Norihiro Tanahashi\footnote[3]{tanahasi@yukawa.kyoto-u.ac.jp}
      }}\\
    \vspace{0.8cm}
    
   ${}^1${\it DAMTP, University of Cambridge, Centre for Mathematical Sciences,}\\
{\it Wilberforce Road, Cambridge, CB3 0WA, United Kingdom}\\
   ${}^{2, 3}${\it Yukawa Institute for Theoretical Physics, Kyoto University,
   Kyoto 606-8502, Japan} 

   \vspace{1cm}
   {\large \today}
  \end{center}

  \vskip 3em

 \begin{abstract}
  We study the non-equilibrium condensation process in a holographic
  superconductor.
  When the temperature $T$ is smaller than a critical
  temperature $T_c$,
  there are two black hole solutions,
  the Reissner-Nordstr\"{o}m-AdS black hole
  and a black hole with a scalar hair. 
  In the boundary theory, 
  they can be regarded as the supercooled normal phase and the superconducting
  phase, respectively.
  We consider perturbations 
  on supercooled Reissner-Nordstr\"{o}m-AdS black holes and 
  study their non-linear time evolution
to know about physical phenomena associated with rapidly-cooled superconductors.
  We find that, for $T<T_c$, 
  the initial perturbations grow exponentially and, eventually, 
  spacetimes approach the hairy black holes.
We also clarify how the relaxation process from a far-from-equilibrium state 
proceeds in the boundary theory 
by observing the time dependence of the superconducting order parameter.
  Finally, we study the time evolution of event and apparent horizons and 
  discuss their correspondence with the entropy of the boundary theory.
Our result gives a first step toward the holographic understanding of the 
non-equilibrium process in superconductors.
\end{abstract}

\end{titlepage}

\section{Introduction}

Over the past decade the AdS/CFT correspondence has received increasing
attention and many studies have been made on it.
It has been believed that the AdS/CFT correspondence plays a central role in
the study of the strongly coupled region of quantum field theory 
because it is
simply described by classical gravity theory on AdS spacetimes~\cite{Ma,GKP,Wi}.
Recently, the duality between the superconductor
and gravity theory has been
proposed~\cite{Gubser:2008px,Hartnoll:2008vx,Hartnoll:2008kx}
as a new application of the AdS/CFT correspondence, in which
the simplest gravity theory is given by 
Einstein-Maxwell-charged scalar theory with negative cosmological
constant.
In this gravity theory 
one of static solutions is 
the well-known Reissner-Nordstr\"{o}m-AdS black hole solution, in which
the scalar field vanishes.
It is known that this solution is unstable when the temperature $T$ is lower
than a critical temperature $T_c$. 
In consequence of the instability, it is expected that the scalar field
will condense into a non-vanishing profile and eventually the black hole
will have the scalar hair breaking the $U(1)$-gauge symmetry spontaneously.
For low temperature $T<T_c$, such a static solution 
has been constructed numerically.
It was shown that the solution has similar properties with
superconductors~\cite{Hartnoll:2008kx}.
Thus, the instability of the Reissner-Nordstr\"{o}m-AdS black hole for
low temperature was identified with the superconducting phase transition.
In the same way,
the Reissner-Nordstr\"{o}m-AdS and hairy
black holes in the gravitational theory 
were identified with the normal and
superconducting phases 
of a superconductor realized within 
the dual field theory, respectively.
%
Such holographic superconductors are considered as a hopeful approach
to understand the property of strongly correlated electron systems.


The non-equilibrium process of strongly correlated systems such as
superconductors is not fully understood 
because of difficulties in its theoretical treatment,
and has been attracting much 
attention~\cite{Mondello:1990zz,Liu:1991zzc,Mondello:1992zz,Korutcheva:1998,Stephens:2001fv}.
The AdS/CFT correspondence offers a novel approach to this longstanding problem.
To understand non-equilibrium process of strongly correlated systems, we
should simply solve classical dynamics of gravitational systems in the
bulk thanks to the duality.
Many attempts in this direction have been done to get insights for
strongly coupled quark-gluon plasma, relativistic hydrodynamics and so
on~\cite{Janik:2005zt,Kovtun:2004de,Bhattacharyya:2008jc,Grumiller:2008va,Gubser:2008pc,AlvarezGaume:2008fx,Lin:2009pn,Gubser:2009sx,Bhattacharyya:2009uu,Chesler:2009cy}. 
For near equilibrium dynamics of superconductor, 
there are several approaches from the holography.
For example, 
transport coefficients, such as the electric conductivity, 
were obtained by the linear response theory~\cite{Hartnoll:2008vx,Hartnoll:2008kx,Amado:2009ts}.
(See also~\cite{Hartnoll:2009sz,Horowitz:2010gk} and references
therein.)
The static and dynamic critical phenomena were also studied
in~\cite{Maeda:2009wv}.\footnote{
In addition, the superfluid hydrodynamics of this system has been
discussed in~\cite{Herzog:2008he,Herzog:2009md}.}
These studies played important rolls in understanding the 
holographic superconductor.
However, very little progress has been made in the regime that the theory is 
far from equilibrium.
In this paper, we give a holographic approach to understand the 
non-equilibrium process of superconductors.

As we mentioned before, for low temperature $T<T_c$ there are two phases of black holes in
the gravity side,
the Reissner-Nordstr\"{o}m-AdS 
and the hairy black holes, 
and they are regarded as supercooled normal phase and superconducting
phases, respectively.
In this paper, we consider small perturbations on the supercooled
Reissner-Nordstr\"{o}m-AdS black holes and study non-linear time
evolution of them.
Because of the instability of the Reissner-Nordstr\"{o}m-AdS black holes, 
the initial perturbations will grow exponentially at the beginning of the time evolution.
It is expected that the exponential growth is saturated due to the non-linear effect and 
the spacetimes will approach static solutions.
It is believed that the final states of the time evolution are the hairy black holes
obtained in~\cite{Hartnoll:2008kx}, but there is no proof.\footnote{%
Recently, 
evidence of this conjecture has been given in~\cite{Maeda:2010hf}
in the near critical temperature regime.}
Studying non-linear time evolutions of the Einstein-Maxwell-charged
scalar system, we will show that the final states 
are given by the hairy black holes.
We are also interested in dynamics of the phase transition on the
boundary theory.
To reveal the non-equilibrium process, 
we observe the time dependence of
the superconducting order parameter and study how the normal phase goes
to superconducting phase in the boundary theory in the middle of the time evolution.
We also evaluate the relaxation time scale of
the boundary theory.
Finally, we study the time evolution of event and apparent horizons and 
discuss the correspondence with the entropy of the boundary theory.

The organization of this paper is as follows.
In Section~\ref{sec:EOM}, 
we introduce the gravity theory of the holographic superconductor and give equations of motion.
Assuming the plane symmetry, we obtain the (1+1)-dimensional partial
differential equations (PDEs). We also give the outline of our
calculations.
In Section~\ref{sec:Asym},
we solve the equations of motion near the AdS boundary 
in order to obtain boundary conditions at the AdS boundary.
In Section~\ref{sec:method},
we explain the numerical method to solve the equations of motion.
In Section~\ref{sec:result},
we give the numerical results of non-linear time evolution.
We find that the final states of the time evolutions coincide with the hairy
black holes for general initial conditions. 
In the middle of time evolution, we study the time dependence of
superconducting order parameters and measure the relaxation time of them.
The time evolutions of event and apparent horizons are also studied.
The final section is devoted to conclusions.


\section{Equations of motion}\label{sec:EOM}

We consider the $4$-dimensional Einstein-Maxwell-charged scalar theory 
with negative cosmological constant, whose action is given by 
\begin{equation}
S=\int d^4 x \sqrt{-g}\left[
R + \frac{6}{L^2} -\frac{1}{4}F_{\mu\nu}F^{\mu\nu}-|\partial_\mu \psi
-i q A_\mu\psi|^2 -m^2|\psi|^2
\right]\ ,
\label{SCaction}
\end{equation}
where $L$ is the AdS curvature scale and $q$ is the 
$U(1)$-charge of the complex scalar field $\psi$.
The field strength is defined by $F=dA$ as usual.
This theory is introduced
in~\cite{Gubser:2008px,Hartnoll:2008vx,Hartnoll:2008kx}
as a gravity dual of a superconductor.
In addition to the diffeomorphism symmetry, this action has
the local $U(1)$ symmetry,
\begin{equation}
 A\to A+d\lambda\ ,\quad \psi\to e^{iq\lambda}\psi\ ,
\end{equation} 
where $\lambda$ is an arbitrary scalar function.
Hereafter, we take the unit of $L=1$.

For simplicity, we assume that the spacetime has the plane symmetry.
Using the diffeomorphism and $U(1)$ gauge symmetries together with the
assumed plane symmetry, we can take the metric ansatz without loss of generality as
\begin{equation}
\begin{split}
&ds^2=-\frac{1}{z^2}\left[F(t,z)dt^2 +2dtdz\right] + \Phi(t,z)^2(dx^2+dy^2)\ ,\\
&A=\alpha(t,z) dt\ ,\\
&\psi=\psi(t,z)\ . 
\end{split}
\label{plane_sym}
\end{equation}
We use the ingoing Eddington-Finkelstein coordinates, $(t,z)$.
In these coordinates, the AdS boundary is located at $z=0$.
These coordinates are convenient for our numerical calculations
since we can easily extend time slices defined by constant-$t$ surfaces 
into the inside of the event horizon 
and also set the 
AdS boundary to be a constant-$z$ plane.
Note that diffeomorphism and $U(1)$ gauge symmetries have not been completely
fixed, because the form of the variables~(\ref{plane_sym}) is invariant under the residual
symmetry, $1/z\to 1/z+g(t)$, $\alpha\to \alpha + \partial_t \lambda(t)$
and $\psi\to e^{iq\lambda(t)}\psi$. 
These residual gauge symmetries will be fixed by the boundary conditions.
The complete set of the equations of motion are given by
\begin{align}
&(\Phi D\Phi)' - \frac{\Phi^2}{4z^2}\left(
\frac{1}{2}z^4 \alpha'^2 + m^2|\psi|^2 - 6
\right)=0\ ,
\label{DEV1}\\
&
2z^2(D\psi)'+iqz^2\alpha' \psi 
+2 z^2 \Phi^{-1} (D\Phi) \psi'
+2 z^2 \Phi^{-1} \Phi' D\psi 
+m^2\psi=0\ ,
\label{DEV2}\\
&
(z^2(z^{-2}F)')'
-z^2\alpha'{}^2 +4\Phi^{-2}(D\Phi)\Phi'
-(\psi^\ast{}' D\psi + \psi' D\psi^\ast)=0\ ,
\label{DEV4}\\
&
 2z^2(D\alpha)' + z^4 (z^{-2}F)'\alpha' + 4z^2\Phi^{-1}(D\Phi)\alpha'
-2iq(\psi D\psi^\ast-\psi^\ast D\psi)=0
\ ,\label{DEV5}
\end{align}
and
\begin{align}
&
\Phi^{-2}C_1\equiv 
-2\Phi^{-1}D^2\Phi-\left(F'-\frac{2}{z}F\right)\Phi^{-1}D\Phi
 -|D\psi|^2=0\ ,
\label{CON1}\\
&
\Phi^{-2}C_2\equiv -2z^3\Phi^{-1}(z \Phi''+2\Phi')-z^4|\psi'|^2=0\ ,
\label{CON2}\\
&
\Phi^{-2}C_3\equiv -z^2[
z^2\alpha''+2z\alpha'+2z^2\Phi^{-1}\Phi'\alpha' +iq(\psi \psi^\ast{}'-\psi^\ast \psi')
]=0\ ,
\label{CON3}
\end{align}
where $'\equiv \partial_z$ and derivative operator $D$ is defined as
\begin{equation}
\begin{split}
&D \Phi = \partial_t \Phi - F \partial_z \Phi/2\ ,\quad
D^2 \Phi=\partial_t (D\Phi) - F \partial_z (D\Phi)/2\ ,\quad
D \alpha = \partial_t \alpha - F \partial_z \alpha/2\ ,\\
&D \psi = \partial_t \psi - F \partial_z \psi/2 -iq\alpha \psi\ ,\quad
D \psi^\ast = \partial_t \psi^\ast - F \partial_z \psi^\ast/2 +iq\alpha \psi^\ast\ .
\end{split}
\label{dotdef}
\end{equation}
Here, the operator $\partial_t - F \partial_z /2$ represents the 
derivative along the outgoing null vector.
We regard Eqs.~(\ref{DEV1}-\ref{DEV5}) as evolution equations and 
Eqs.~(\ref{CON1}-\ref{CON3}) as constraint equations.
The constraint equations satisfy
\begin{align}
 &\partial_z C_1=0\ ,\label{eqC1}\\
&D C_2=z^2\partial_z (F/z^2)C_2 +\frac{1}{2}z^2(\partial_z \alpha)C_3\ ,\label{eqC2}\\
&
 D C_3=\frac{1}{2}z^2\partial_z (F/z^2)C_3 \ ,\label{eqC3}
\end{align}
where $D{C}_i=\partial_tC_i - F \partial_z C_i/2$ $(i=2,3)$.
To derive the above equations, we have used evolution
equations~(\ref{DEV1}-\ref{DEV5}).
As we can see from Eqs.~(\ref{eqC1}-\ref{eqC3}), 
the evolution equations guarantee that the constraint equations to be satisfied 
if $C_1=0$ and $C_2=C_3=0$ are 
satisfied on the AdS boundary ($z=0$) and on the initial surface ($t=0$), respectively.
Therefore, we solve the evolution equations for the time evolution and 
use 
the constraint equations only at the AdS boundary and the initial
surface to give the boundary conditions for the time evolution.

One of static solutions of the equations of motion is merely
the Reissner-Nordstr\"{o}m-AdS black hole solution (with planar horizon), which is given by
\begin{equation}
 F=1-2M z^3+\frac{1}{4}Q^2z^4\ ,\quad
\Phi=\frac{1}{z}\ ,\quad
\alpha=Qz\ ,\quad
\psi=0\ .
\end{equation}
Parameters $M$ and $Q$ are proportional to mass and charge of the black
hole, respectively.
For this solution, the complex scalar field has a trivial configuration
$\psi = 0$.
That is, this black hole has no ``hair'' except for the mass and electric charge. 
The event horizon is located at $z=z_+$ determined by $F(z_+)=0$. 
In terms of the
horizon radius $z_+$, we can rewrite $M$ as
\begin{equation}
 M=\frac{1}{2z_+^3}\left(1+\frac{1}{4}Q^2z_+^4\right)\ .
\label{Mzp}
\end{equation}
The temperature of the black hole is given by
\begin{equation}
 T=\left.-\frac{1}{4\pi}\frac{dF}{dz}\right|_{z=z_+}
=\frac{12-z_+^{4}Q^2}{16\pi z_+}\ .
\label{tempe}
\end{equation}
It is known that this solution is unstable when the temperature $T$ is smaller
than a critical temperature $T_c$. The numerical value of the $T_c$ is
obtained in~\cite{Gubser:2008px,Hartnoll:2008vx,Hartnoll:2008kx}.
In consequence of the instability, it is expected that the scalar field
will grow and eventually the black hole
will have the scalar hair.
Thus, the $U(1)$-gauge symmetry is spontaneously broken due to
condensation of the complex scalar field.
Indeed, for low temperature $T<T_c$ at fixed charge $Q$, static
solutions of the hairy black hole with $\psi\ne 0$ exist and have been constructed numerically~\cite{Hartnoll:2008kx}.
Similarly to the previous case, the temperature of the hairy black hole is given by
\begin{equation}
 T=\left.-\frac{1}{4\pi}\frac{dF}{dz}\right|_{z=z_+},
\end{equation}
where $F(z_+)=0$. 

\begin{figure}
  \begin{center}
   \includegraphics[width=.48\linewidth,angle=270,clip]{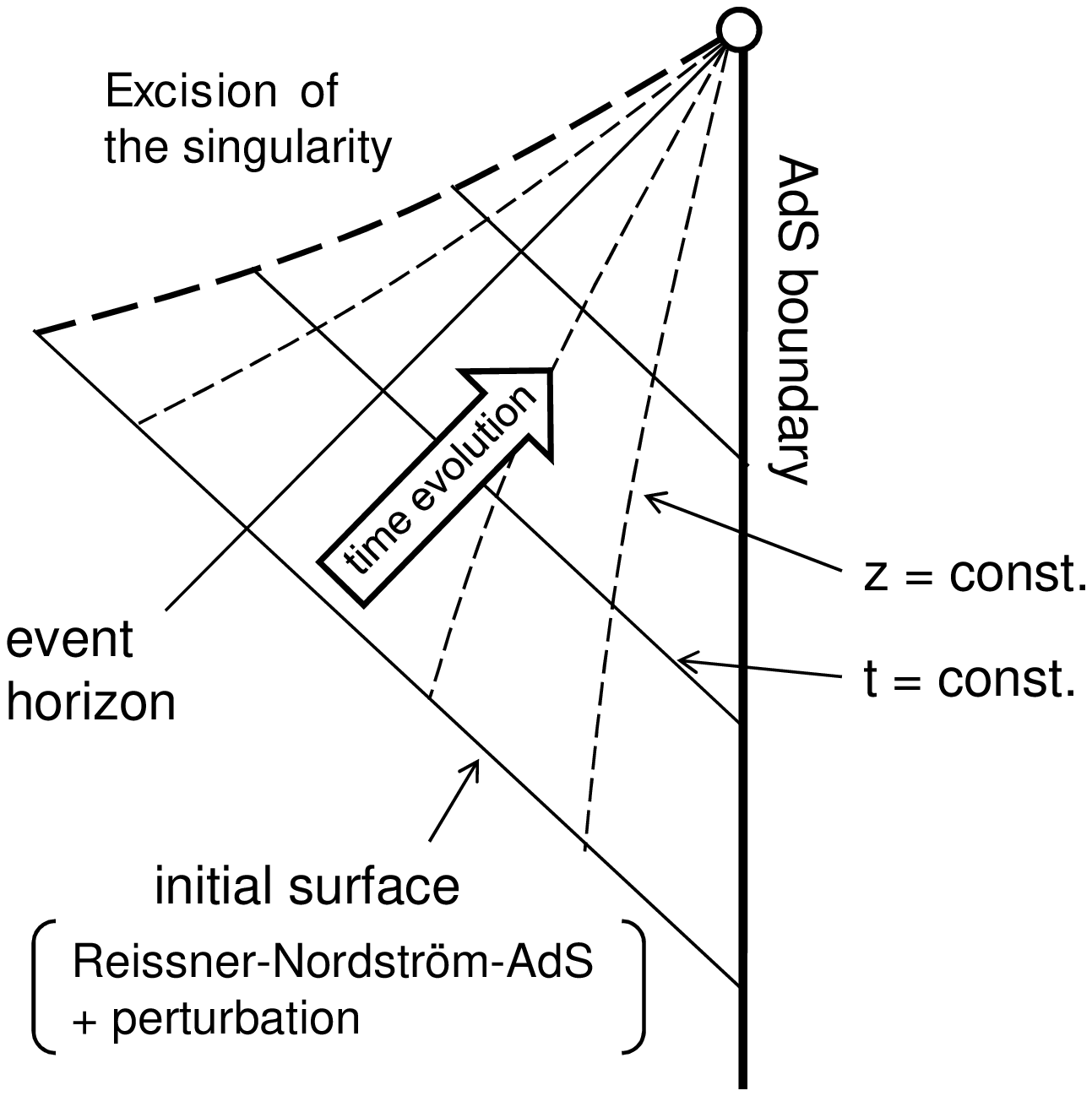}
   \caption{
\label{fig:ponchi}
Schematic of our setting. We take the ingoing
   Eddington-Finkelstein coordinates, in which time slices are given by
   null surfaces. We consider a small perturbation on the 
Reissner-Nordstr\"{o}m-AdS spacetime as initial data on an initial
   surface and study its non-linear time evolution.}
  \end{center}
\end{figure}

This instability was identified with the superconducting phase
transition in the dual theory~\cite{Gubser:2008px,Hartnoll:2008vx,Hartnoll:2008kx}.   
In this paper, we will investigate the dynamical process of the superconducting phase
transition in the view of the gravity theory.
We show the schematic of our setting in Figure~\ref{fig:ponchi}.
Since we are taking the ingoing Eddington-Finkelstein coordinates~(\ref{plane_sym}),
our time slices are given by null surfaces.
We consider 
a slightly-perturbed Reissner-Nordstr\"{o}m-AdS spacetime as initial data,
and study non-linear time evolution from it.
At the AdS boundary, we give appropriate boundary conditions, as we explain in 
the subsequent section.
Inside the event horizon, we excise a region before encountering the singularity for the numerical
calculation. The detailed procedure will be explained in following
sections.

\section{Asymptotic expansion}\label{sec:Asym}
Now, we determine the asymptotic form of the variables at the AdS
boundary $z=0$ to clarify the boundary conditions for the time evolution.
Hereafter, we set the mass of the complex scalar field  as $m^2 = -2L^{-2}$
following~\cite{Hartnoll:2008vx,Hartnoll:2008kx}. Then, 
we can expand variables as
\begin{equation}
\begin{split}
&F(t,z)=1 + F_1(t)z + F_2(t)z^2 + F_3(t)z^3 + \cdots\ ,\\
&\Phi(t,z)=1/z +\Phi_{0}(t) + \Phi_1(t)z + \Phi_2(t)z^2 + \Phi_3(t)z^3 + \cdots\ ,\\
&\alpha(t,r)=\alpha_0(t) +\alpha_1(t)z + \alpha_2(t)z^2 + \alpha_3(t)z^3 +\cdots\ ,\\
&\psi(t,r)=\psi_1(t)z + \psi_2(t)z^2 + \psi_3(t)z^3 +\cdots\ .
\end{split}
\label{asym}
\end{equation}
Using residual gauge symmetries $1/z\to 1/z+g(t)$ and 
$\alpha\to\alpha+\partial_t\lambda(t)$, 
we put $\Phi_0(t)=0$ and $\alpha_0(t)=0$.
Substituting Eq.~(\ref{asym}) into Eqs.~(\ref{DEV1}-\ref{CON3})
and solving the equations order by order,
we obtain the expansion coefficients as
\begin{equation}
\begin{split}
&F_1=0\ ,\quad 
F_2 =-\frac{1}{2}\psi_1\psi_1^\ast \ ,\quad
\dot{F}_3= \frac{1}{2}\big[
\psi^\ast_1 \ddot{\psi_1}+ \psi_1 \ddot{\psi}^\ast_1
- \psi_1 \dot{\psi}^\ast_2 
- \psi^\ast_1 \dot{\psi}_2\big]
\\
&\Phi_1=-\frac{1}{4}|\psi_1|^2\ ,\quad
\Phi_2=-\frac{1}{6}(\psi_1\psi_2^\ast+\psi_1^\ast\psi_2)\ ,
\\
&\Phi_3=
-\frac{1}{6} |\psi_2|^2 
-\frac{11}{96} |\psi_1|^4
-\frac{1}{8} (\psi_1 \dot{\psi}^\ast_2+\psi^\ast_1 \dot{\psi}_2)
\\
&\dot{\alpha}_1=-iq (-\psi_2\psi^\ast_1 +\psi_1\psi^\ast_2 +
 \dot{\psi}_1\psi^\ast_1-\psi_1 \dot{\psi}^\ast_1)\ ,\quad
\alpha_2 =-\frac{1}{2}iq ( -\psi_2\psi^\ast_1 +\psi_1\psi^\ast_2) \ ,\\
%
&\psi_3=-\frac{1}{2}iq\alpha_1\psi_1+\dot{\psi}_2+\frac{1}{2}\psi_1^2\psi_1^\ast\ ,
\end{split}
\label{coeff}
\end{equation}
where ${}^\cdot \equiv d/dt$.
We find that all the coefficients $F_i(t)$, $\Phi_i(t)$, $\alpha_i(t)$ and
$\psi_i(t)$ can be expressed by $\psi_1(t)$ and $\psi_2(t)$.
This means that if the two functions $\psi_1(t)$ and $\psi_2(t)$ are given at
the boundary we can determine the bulk in principle.
It is noticed that these are order parameters characterizing normal and
superconducting phases in the boundary theory~\cite{Hartnoll:2008vx,Hartnoll:2008kx}.
In our calculation, we do not give $\psi_2(t)$ at $z=0$, and
$\psi_2(t)$ is determined as a result of the time evolution.
As for the inner boundary, a constant-$z$ surface, no boundary condition is 
required there since it will be a spacelike surface as long as we take 
it behind the horizon.

Note that initial values of $F_3(t)$ and $\alpha_1(t)$ are 
not determined by the asymptotic expansion, because 
the asymptotic expansion gives just time derivative of 
$F_3(t)$ and $\alpha_1(t)$.
In fact, $F_3(t)$ and $\alpha_1(t)$ are related to with mass and
charge of the system. 
Hence initial values of $F_3(t)$ and $\alpha_1(t)$ represent the initial mass
and charge of black holes. We put the initial mass and charge as $M_0$ and
$Q_0$, respectively and, then, $F_3$ and $\alpha_1$ are written as
\begin{equation}
\begin{aligned}
F_3 \equiv & -2M(t)\\
= & -2M_0 + \frac{1}{2}\int^t_0 dt \big[
 \psi^\ast_1 \ddot{\psi_1}+ \psi_1 \ddot{\psi}^\ast_1
+ \dot{\psi}_1 \psi^\ast_2 
+ \dot{\psi}^\ast_1 \psi_2
\big]
-\frac{1}{2}\bigg[\psi_1 \psi^\ast_2+ \psi^\ast_1 \psi_2\bigg]^{t}_0
\ ,\\
\alpha_1 \equiv & Q(t)\\
= & Q_0-iq\int^t_0 dt \big[
-\psi_2\psi^\ast_1 +\psi_1\psi^\ast_2 +
 \dot{\psi}_1\psi^\ast_1-\psi_1 \dot{\psi}^\ast_1
\big] \ ,
\end{aligned}
\label{a1F3}
\end{equation}
where $M(t)$ and $Q(t)$ denote the total mass and charge, respectively.

\section{A method to solve equations of motion}\label{sec:method}

In this section, we explain a method to solve the equations of motion
numerically\footnote{%
Similar calculations were performed in~\cite{Chesler:2009cy} 
in the context of the AdS/QGP.}.
From Eq.~(\ref{coeff}), 
we can see that 
the first few expansion coefficients 
in Eq.~(\ref{asym}) 
do not depend on $\psi_2(t)$ but only on $\psi_1(t)$.
In our calculation, we give $\psi_1(t)$ as a boundary condition at $z=0$.
Moreover, some variables have divergent parts whose form is fully determined by 
the asymptotically-AdS boundary conditions.
Thus, it is reasonable for numerical calculations 
to define new regular variables $\tilde\Phi$, $\tilde\psi$
and $\tilde{F}$ as
\begin{equation}
\begin{split}
&\Phi=1/z + z\Phi_1(t) + z^2\tilde{\Phi}(t,z)\ ,\\
&\psi=z\psi_1(t) + z^2\tilde{\psi}(t,z)\ ,\\
&F=1+zF_1(t)+z^2 F_2(t) + z^3 \tilde{F}(t,z)\ ,
\end{split}
\label{tilde}
\end{equation}
where $\Phi_1(t)$, $F_1(t)$ and $F_2(t)$ are given by Eq.~(\ref{coeff}) 
and they do not depend on $\psi_2(t)$ but only on $\psi_1(t)$.
Note that by these definitions we have $\psi_2(t) = \tilde\psi(t,z)|_{z=0}$. 
We also define $\tilde{D\Phi}$ and $\tilde{D\psi}$ as
\begin{equation}
\begin{split}
&D\Phi=1/(2z^2) + \Phi_1(t)/2 + F_2(t)/2 + z \tilde{D\Phi}(t,z)\ ,\\
&D\psi=-\psi_1(t)/2 + z\tilde{D\psi}(t,z)\ .
\end{split}
\label{dottilde}
\end{equation}
From Eq.~(\ref{dotdef}), $\tilde{D\Phi}$ and $\tilde{D\psi}$ can be
written in terms of $\tilde\Phi$ and  $\tilde\psi$ as
\begin{align}
&2\tilde{D\Phi}=2\dot\Phi_1 - 2\tilde\Phi + \tilde F
+z(2\partial_t \tilde\Phi -\partial_z \tilde\Phi -F_2 \Phi_1 ) 
\notag\\
&\hspace{3cm}
-z^2(2F_2  \tilde\Phi+\Phi_1 \tilde F)
-z^3(F_2 \partial_z\tilde\Phi + 2\tilde F \tilde\Phi)
-z^4 \tilde F \partial_z \tilde\Phi\ ,
\\
&2\tilde{D\psi}=
2\dot\psi_1  -2\tilde{\psi} + z(2\partial_t \tilde{\psi} -\partial_z
 \tilde{\psi} -F_2 \psi_1 ) 
\notag\\
&\hspace{3cm}
-z^2(2F_2 \tilde\psi + \psi_1 \tilde F) 
- z^3(F_2 \partial_z \tilde\psi+2\tilde\psi \tilde F)
-z^4 \tilde F \partial_z \tilde\psi\ .
\end{align}
In terms of $\tilde\Phi$, $\tilde\psi$, $\tilde{F}$, $\tilde\Phi$,
$\tilde\psi$, $\tilde{D\Phi}$ and $\tilde{D\psi}$, we can rewrite
evolution equations~(\ref{DEV1}-\ref{DEV5}) as
\begin{align}
&(\tilde{D\Phi})' + X_1[\tilde\Phi]\tilde{D\Phi}
 +X_2[\tilde\Phi,\tilde\psi,\alpha]=0\ ,
\label{DEV12}\\
&
(\tilde{D\psi})'+Y[\tilde{D\Phi},\tilde\Phi,\tilde\psi,\alpha]=0\ ,
\label{DEV22}\\
&
(z^2\tilde{F})'' +
 Z[\tilde{D\Phi},\tilde{D\psi},\tilde\Phi,\tilde\psi,\alpha]=0\ ,
\label{DEV42}\\
&
 (D\alpha)' +
 W[\tilde{D\Phi},\tilde{D\psi},\tilde\Phi,\tilde\psi,\alpha,\tilde{F}]=0
\ ,\label{DEV52}
\end{align}
and constraint equations~(\ref{CON2}) and (\ref{CON3}) as
\footnote{
We do not use Eq.~(\ref{CON1}),
one of the constraint equations, hereafter.
This equation has been already used to derive the asymptotic form of
the variables~(\ref{coeff}).
In our calculations, 
Eq.~(\ref{CON1}) will be guaranteed
to be satisfied by the 
boundary conditions~(\ref{coeff}). 
}
\begin{align}
&\tilde{\Phi}''+P_1[\tilde\Phi]\tilde\Phi' +
 P_2[\tilde\Phi,\tilde\psi]=0\ ,
\label{CON22}\\
&
\alpha''+Q_1[\tilde\Phi]\alpha' + Q_2[\tilde\psi]=0\ .
\label{CON32}
\end{align}
where $X_1$, $X_2$, $Y$, $Z$, $W$ $P_1$, $P_2$, $Q_1$ and $Q_2$ are
functions of each arguments.
Since the expression of these functions are tedious, we do not write their
explicit expressions.

The procedure to solve Eqs.~(\ref{DEV12}-\ref{CON32}) as follows.
To begin with, we must construct initial data of $\tilde\Phi(t=0,z)$,
$\tilde\psi(t=0,z)$ and $\alpha(t=0,z)$ on the initial surface $t=0$
that satisfy the constraint equations~(\ref{CON22}) and (\ref{CON32}).

\begin{enumerate}
\item 
On the initial surface $t=0$,
we prepare initial data of $\tilde\psi(t=0,z)$, initial mass $M_0$ and
      charge $Q_0$.
\item
Integrating Eq.~(\ref{CON22}) and (\ref{CON32}) from $z=0$ to $z=z_0$, 
we can obtain initial data $\tilde\Phi(t=0,z)$ and $\alpha(t=0,z)$. 
Here, $z=z_0$ is the boundary of computational domain, which we 
set to be between the horizon and the black hole singularity.
For these radial integrations, we need boundary conditions 
$\tilde\Phi(t=0,z)|_{z=0}$ and $\alpha(t=0,z)|_{z=0}$.
They are given by the asymptotic
expansion~(\ref{asym}) as
\begin{equation}
\begin{split}
&
\tilde\Phi(t=0,z) \big|_{z=0}=\Phi_2(0)\ ,\quad
\partial_z \tilde\Phi(t=0,z) \big|_{z=0}=\Phi_3(0)\ ,\\
&
\alpha(t=0,z) \big|_{z=0} = 0\ , \quad
\partial_z \alpha(t=0,z) \big|_{z=0} = \alpha_1(0)\ ,
\end{split}
\end{equation}
where coefficients $\Phi_2(0)$, $\Phi_3(0)$ and $\alpha_1(0)$ are written
by $\psi_1(0)$ and $\psi_2(0)$ as shown in Eq.~(\ref{coeff}).
The $\psi_2(0)$ can be read off from $\tilde\psi(t=0,z)|_{z=0}$.
\end{enumerate}

Once we have obtained the data of $\tilde\Phi(t,z)$, $\tilde\psi(t,z)$ and
$\alpha(t,z)$ on a constant-$t$ surface at an arbitrary time $t$, we can
calculate the data of 
$\tilde\Phi(t+\delta t,z)$, $\tilde\psi(t+\delta t,z)$ and
$\alpha(t+\delta t,z)$ on the next time slice $t+\delta t$ as follows.

\begin{enumerate}[i.]
\item
Integrating Eqs.~(\ref{DEV12}-\ref{DEV52}) from $z=0$ to $z=z_0$, we obtain
the $\tilde{D\Phi}$, $\tilde{D\psi}$, $D\alpha$ and $\tilde{F}$ on the
     constant-$t$ surface, on which $\tilde\Phi(t,z)$,
     $\tilde\psi(t,z)$ and $\alpha(t,z)$ have been known.
Boundary conditions for the radial integration are given by the 
asymptotic expansion~(\ref{asym}) as
\begin{equation}
\begin{split}
&\tilde{D\Phi}(t,z)\big|_{z=0}=
-\Phi_2(t)+\frac{1}{2}F_3(t)-\frac{1}{4}(\psi_1(t)\dot\psi^\ast_1
 (t)+\psi^\ast_1(t)\dot\psi_1 (t))\
 ,\\
&\tilde{D\psi}(t,z)\big|_{z=0}=\dot{\psi}_1(t)-\psi_2(t)\ ,\quad
D\alpha(t,z)\big|_{z=0}=-\frac{1}{2}\alpha_1(t)\\
&z^2\tilde{F}(t,z)\big|_{z=0}=0\ ,\quad
\partial_z [z^2\tilde{F}(t,z)]\big|_{z=0}=0
\ .
\end{split}
\end{equation}
where $\Phi_2(t)$, $F_3(t)$ and $\alpha_1(t)$ are written by $\psi_1(t)$
and $\psi_2(t)$ as in Eq.~(\ref{coeff}) and (\ref{a1F3}).
\item
We calculate $\partial_t \Phi(t,z)$, 
$\partial_t \alpha(t,z)$ and 
$\partial_t \psi(t,z)$ from
Eq.~(\ref{dotdef}).
We use the upwind differencing scheme for advection terms in Eq.~(\ref{dotdef}).
\item
We obtain variables $\tilde\Phi(t+\delta t,z)$, $\tilde\psi(t+\delta t,z)$ and
$\alpha(t+\delta t,z)$ on the next time slice $t+\delta t$ by using time
     derivative of each variables $\partial_t \Phi(t,z)$, 
$\partial_t \alpha(t,z)$ and $\partial_t \psi(t,z)$.
\end{enumerate}
 By repeating the above procedure i $\sim$ iii, we can numerically
 calculate time evolution of the system.

\section{Non-equilibrium Condensation Process}\label{sec:result}

In this section, we summarize the numerical results obtained by the evolution 
scheme of the previous section.
In  Section~\ref{Sec:ID}, we describe the free parameter and the initial conditions.
We summarize the general properties of the bulk field dynamics in Section~\ref{Sec:bulk}
After that, we study the dynamics of the order parameter of the boundary 
theory, which is the boundary value of the bulk scalar field, in Section~\ref{Sec:order}.
We also investigate its growth and decay rates around the initial and final 
states in Section~\ref{Sec:growth}.
In Section~\ref{Sec:horizon}, we focus on the dynamics of the event and the apparent horizons.
If a black hole is stationary, then the entropy of the boundary theory,
which is another important physical quantity, will be straightforwardly
identified with the horizon area in the bulk.
However, it is not so obvious in dynamical cases.
In order to argue their relevance to the boundary theory entropy, 
we clarify to what extent the bulk spacetime is dynamical and give some 
consideration on how to identify the boundary and the bulk spacetime in 
Section~\ref{Sec:non-staticity}.

\subsection{Initial data and parameters}
\label{Sec:ID}

As explained in Section~\ref{sec:method}, 
we should specify the $\tilde\psi(t=0,z)$, $M$ and $Q$
on the initial surface.
We treat $M$ and $Q$ as fixed parameters in this section since they are 
conserved quantities in our setting as we will see later.
Without loss of generality, we can fix one of the parameters using
the scaling symmetry\footnote{%
This is just a residual coordinate transformation which
preserves the form of metric and gauge field in Eq.~(\ref{plane_sym})
as well as 
the boundary conditions $F(t,z)\simeq 1$ and $\Phi(t,z)\simeq 1/z$ for $z\to 0$.
},
\begin{align}
 &(t,z,x,y)\to(kt,kz,kx,ky)\ ,\\
 &F \to F\ ,\quad \Phi \to \Phi/k\ ,\quad \alpha \to \alpha/k\ ,\quad \psi \to \psi\ ,\\
 &M \to M/k^3\ ,\quad Q \to Q/k^2\ ,\quad T \to T/k\ .
\label{scaling}
\end{align}
Using this scaling symmetry, in our numerical calculation, we put
\begin{equation}
M=\frac{1}{2}\left(1+\frac{1}{4}Q^2\right)\ .
\end{equation}
This condition implies that 
the {\it horizon radius} of the initial black hole is set to
unity (see Eq.~(\ref{Mzp}))\footnote{%
To be accurate, the $z_+$ defined by Eq.~(\ref{Mzp}) 
slightly differs from the apparent horizon position for our initial data,
because we will give small perturbations on the
Reissner-Nordstr\"{o}m-AdS solutions.
%
}.
As for the initial data of $\tilde\psi$,
we consider the Gaussian perturbation on the Reissner-Nordstr\"{o}m-AdS
spacetime as
\begin{equation}
 \tilde\psi(t=0,z)= \frac{\mathcal{A}}{\sqrt{2\pi}\,\delta}
\exp\left[
-\frac{(z-z_m)^2}{2\delta^2}
\right]
\label{ini_psi}
\end{equation}
with $\mathcal{A}=0.01$, $\delta=0.05$ and $z_m=0.3$.
We tried several other initial conditions and found that they yield 
qualitatively the same results. Thus, we will show below the results only for the 
initial data given by Eq.~(\ref{ini_psi}).

\subsection{Dynamics of bulk fields}
\label{Sec:bulk}

First of all, we show the dynamics of the bulk scalar field.
As we mentioned before, our numerical calculation is performed under
the boundary condition $\psi_1(t)=0$ at the AdS boundary $z=0$.
This means that we regard $\psi_2(t)$ as the order parameter of
condensation.
Furthermore, it turns out from Eq.~(\ref{a1F3}) that $F_3$ and $\alpha_1$
become time-independent when $\psi_1(t)=0$, namely the total
mass $M$ and the total charge $Q$ are conserved in our calculation.

\begin{figure}[htbp]
  \centering
  \subfigure[$0 \leq tT_c \leq 14$]
  {\includegraphics[height=7.9cm,angle=270,clip]{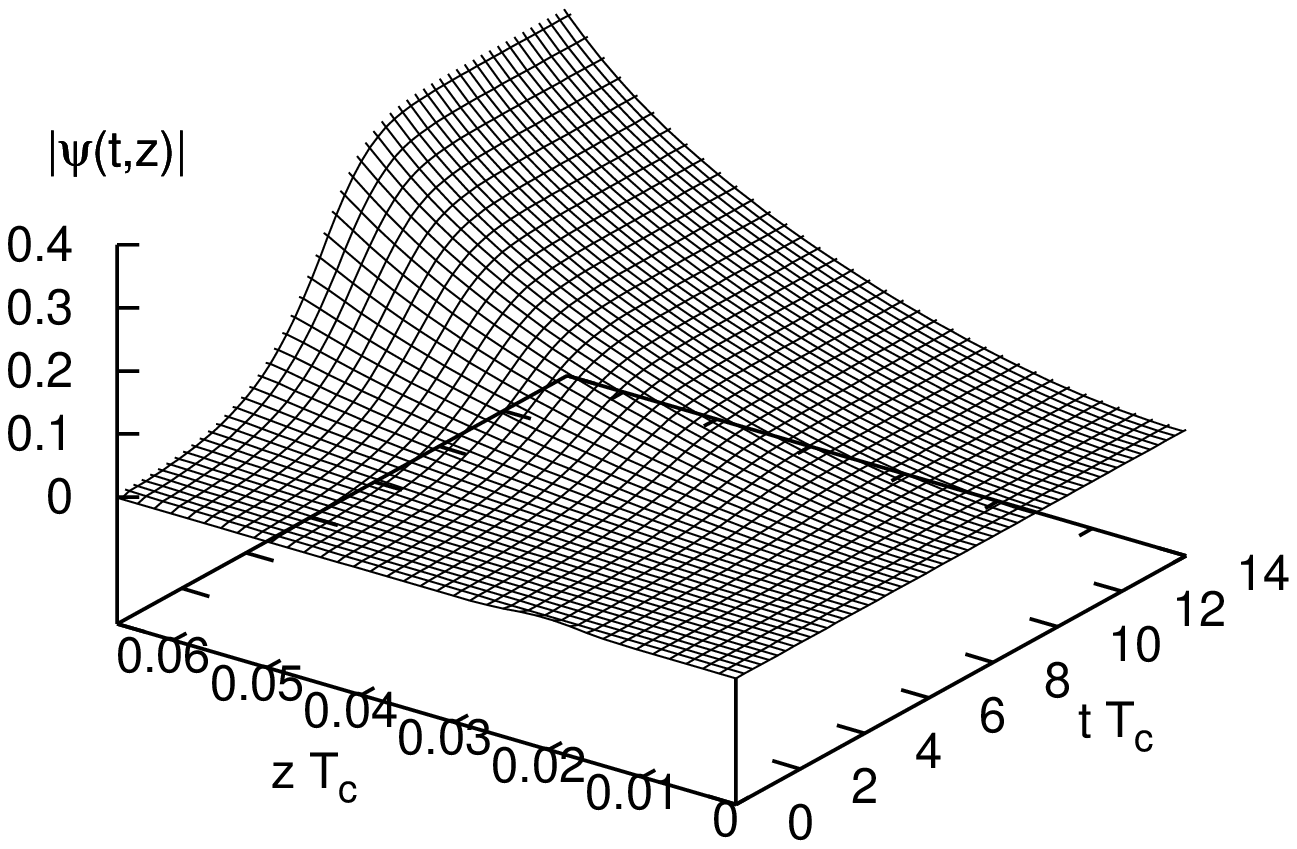}
    \label{fig:psi_global}
  }
  \subfigure[$0 \leq tT_c \leq 0.08$]
  {\includegraphics[height=7.9cm,angle=270,clip]{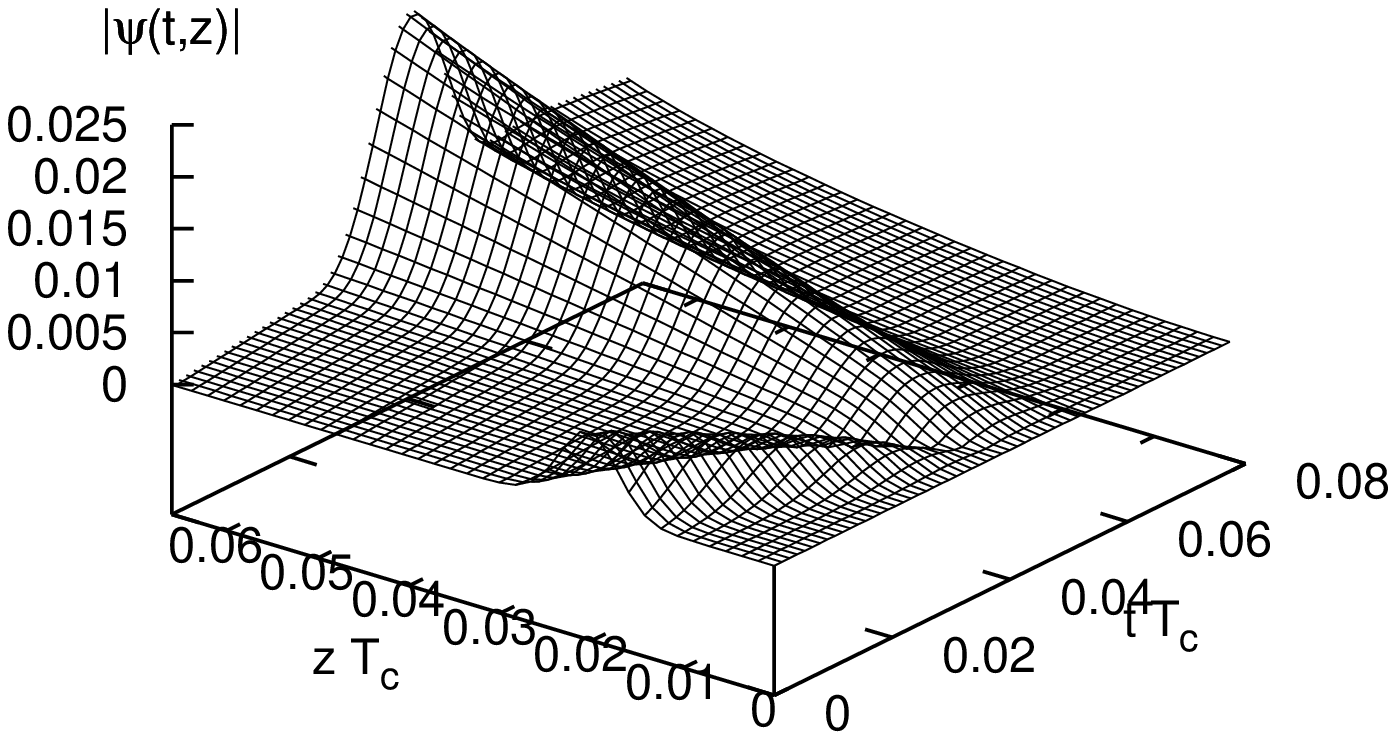} 
    \label{fig:psi_local}
  }
  \caption{\label{fig:psi_sq}
The dynamics of the scalar field for $q=1.0$
   and initial temperature $T/T_c=0.5$.
In Figure~(a), we depict the dynamics of the amplitude of the complex scalar field,
 $|\psi(t,z)|$, on $(t,z)$-plane for $0 \leq tT_c \leq 14$.
Because of the instability of the Reissner-Nordstr\"{o}m-AdS
black hole,  the scalar density grows exponentially for $tT_c\lesssim 6$.
We find that, for $tT_c\gtrsim 6$, the scalar density approaches some
 static function.
In Figure~(b), we depict $|\psi(t,z)|$ for $0 \leq tT_c \leq 0.08$ in
 order to focus on the behavior of the wave packet of the initial perturbation.
We can see that the wave packet is reflected by the AdS boundary at 
$t\simeq 0.04$ and much of the wave packet is absorbed in the black hole
 horizon within $tT_c\lesssim 0.06$.
}
\end{figure}

\begin{figure}[htbp]
  \begin{center}
      \includegraphics[height=8.5cm,angle=270,clip]{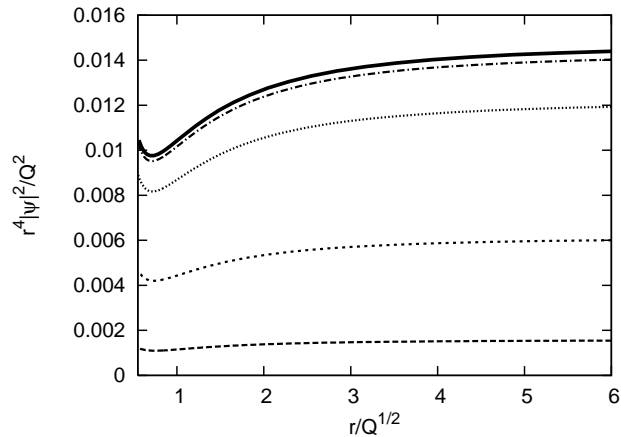}
    \caption{\label{fig:compare}
The function $r^4|\psi|^2/Q^2$ is depicted on fixed
 time slices,
where $r$ is circumference radius defined by $r=\Phi$.
From bottom to top, the curves correspond to 
$tT_c=5.15$, $6.44$, $7.73$ and $9.02$. The top solid curve correspond
 to that of the hairy black hole. 
We can see that the solution approaches the hairy black hole.
}
  \end{center}
\end{figure}

In Figure~\ref{fig:psi_sq}, we depict the dynamics of the amplitude of 
the complex scalar, 
$|\psi(t,z)|$, for $q=1.0$ and $T/T_c=0.5$ at the initial state. 
The critical temperature $T_c$ is evaluated for a fixed charge
$Q$\footnote{%
In this section, 
we investigate the time evolution of the scale invariant variables under
the scaling symmetry~(\ref{scaling}).
For $q=1.0$, $1.5$ and $2.0$, the critical temperature $T_c$ is
respectively given by
$T_c/\sqrt{Q} = 0.03589$, $0.08421$ and $0.1234$.
}.
We can find that much of the wave packet of the initial
perturbation~(\ref{ini_psi}) is instantaneously 
absorbed in the horizon within $tT_c\lesssim 0.06$.
Because of the remnant of the wave packet, which is the unstable mode
contained in the initial perturbation~(\ref{ini_psi}),
the scalar density grows exponentially for $tT_c\lesssim 6$.
The exponential growth is saturated  by the nonlinear effect at
$tT_c \sim 6$. 
In $tT_c\gtrsim 6$, the scalar density approaches a
static solution.
As in Figure~\ref{fig:compare},
we can find
that the static solution coincides with the hairy black hole solution 
obtained in~\cite{Hartnoll:2008kx}. 
Thus, our result gives numerical proof of the conjecture that 
the final state of the instability of the Reissner-Nordstr\"{o}m-AdS
black hole is the hairy black hole.
Our result also 
implies that, for the plane-symmetric perturbations, the hairy
black holes are stable.
It is worth noting that this phase transition from the initial
Reissner-Nordstr\"{o}m-AdS black hole to the final hairy black hole is
a dynamical process under the fixed mass and charge.
It implies that
the temperature of the initial state and that of the final
state are different in general.
Indeed, the temperature of the final hairy black hole increases compared
to the initial temperature due to the phase transition.

\subsection{Dynamics of the order parameter}
\label{Sec:order}

In following subsections, we show some results from the
numerical solutions that are relevant to the dual theory.
In this subsection, we describe the non-linear dynamics of the order parameter of the 
boundary theory.

\begin{figure}[htbp]
  \centering
  \subfigure[$T<T_c$]{\includegraphics[height=8.0cm,angle=270,clip]{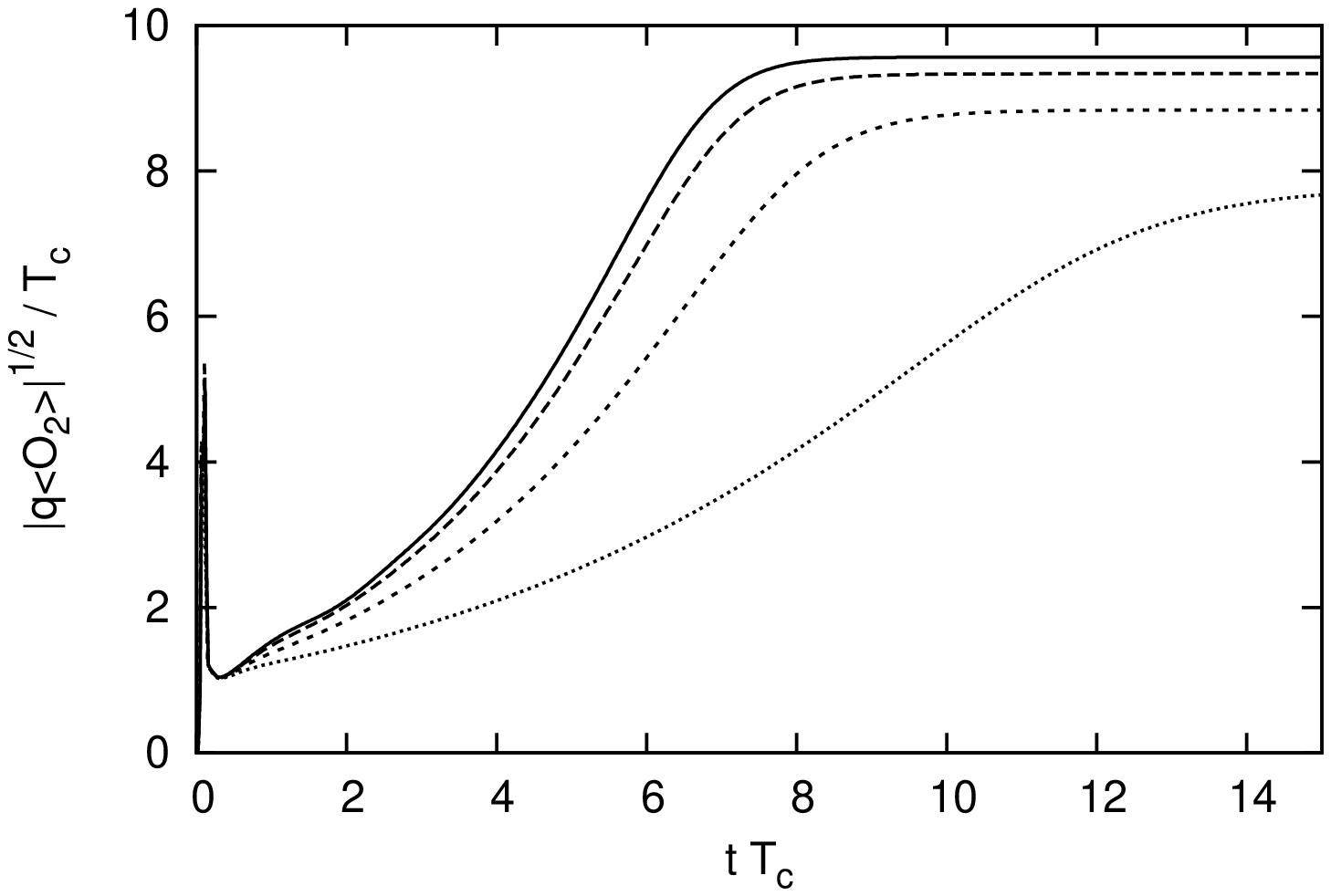}}
  \subfigure[$T>T_c$]{\includegraphics[height=8.0cm,angle=270,clip]{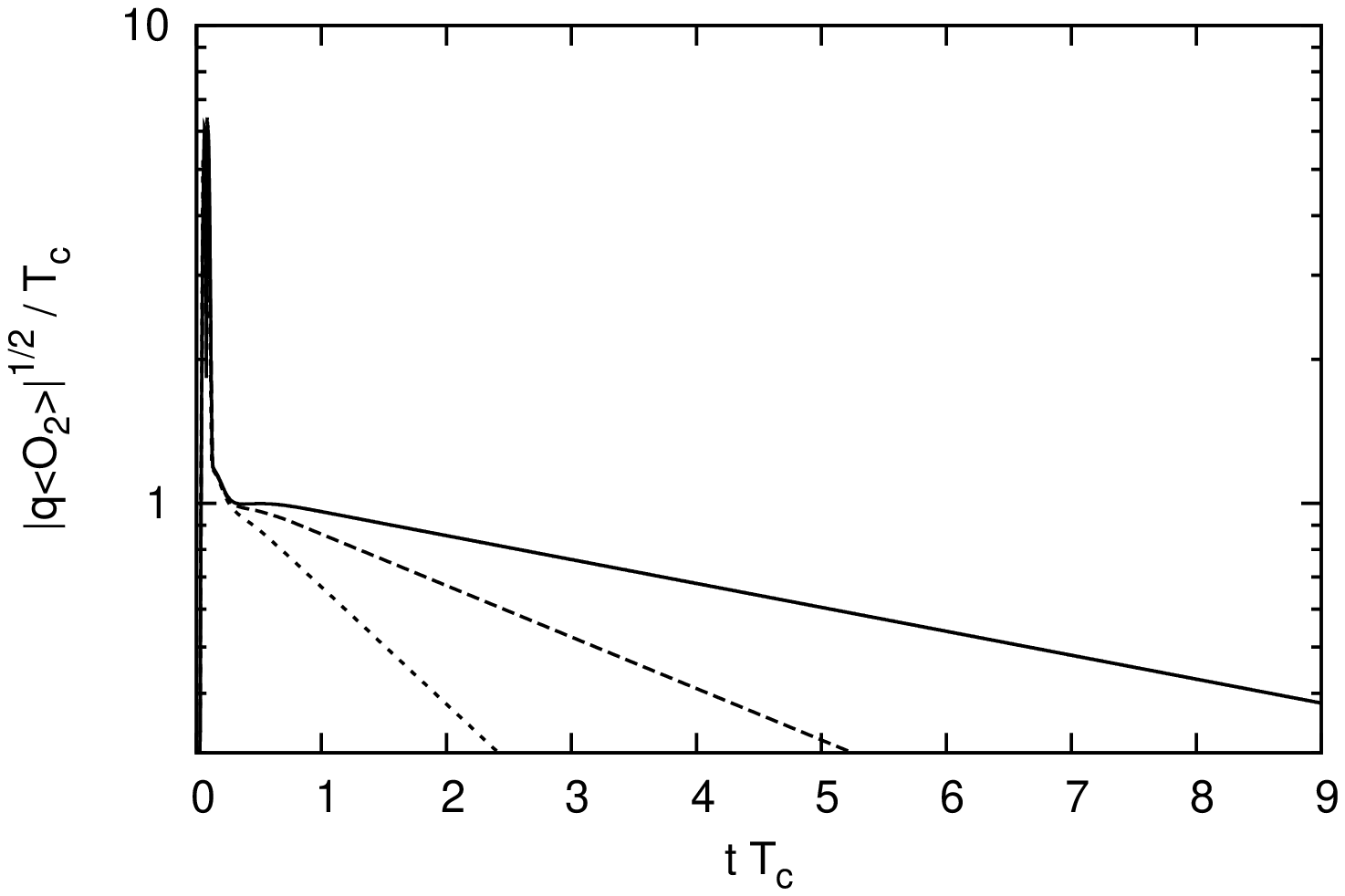}}
  \caption{\label{fig:psi2_dyn}
The dynamics of the order parameter is depicted for $q=1.5$. 
In Figure~(a), the curves from top to bottom correspond to 
$T/T_c=0.2$, $0.4$, $0.6$ and $0.8$. In Figure~(b), the curves from top to 
bottom correspond to 
$T/T_c=1.1$, $1.2$ and $1.4$. 
Note that the vertical axis in Figure~(b) is logarithmic scale,
   while that is linear scale in Figure~(a).}
\end{figure}

From the asymptotic form of the numerical solution $\psi(t,z)$, we can
read off $\psi_2(t)$ defined by Eq.~(\ref{asym}).
The coefficient $\psi_2(t)$ is regarded as the superconducting order
parameter in dual theory. 
Following~\cite{Hartnoll:2008vx,Hartnoll:2008kx}, we define the order
parameter on the boundary theory as
\begin{equation}
 \langle \mathcal{O}_2(t) \rangle \equiv \sqrt{2}\, \psi_2(t)\ .
\end{equation}
In Figure~\ref{fig:psi2_dyn}, we depict the
time dependence of the order parameter 
$(q |\langle \mathcal{O}_2(t)\rangle|)^{1/2}/T_c$, 
which is invariant under the scaling~(\ref{scaling}), for $q=1.5$ and
$T/T_c=0.2$, $0.4$, $0.6$, $0.8$, $1.1$, $1.2$ and $1.4$.
The sharp signal at the small $t$ is caused by the initial perturbation~(\ref{ini_psi}).
For $T>T_c$, the initial perturbation dissipates and the order parameter converges to zero.
On the other hand, for $T<T_c$, the order parameter grows exponentially and
approaches a non-trivial value. 
We find that, for $T<T_c$, 
the more rapidly the order parameter converges to its final value 
for lower temperature.

\subsection{Growth and decay rates}
\label{Sec:growth}

\begin{figure}[htbp]
  \centering
  \subfigure[Normal phase]{\includegraphics[scale=0.52,angle=270]{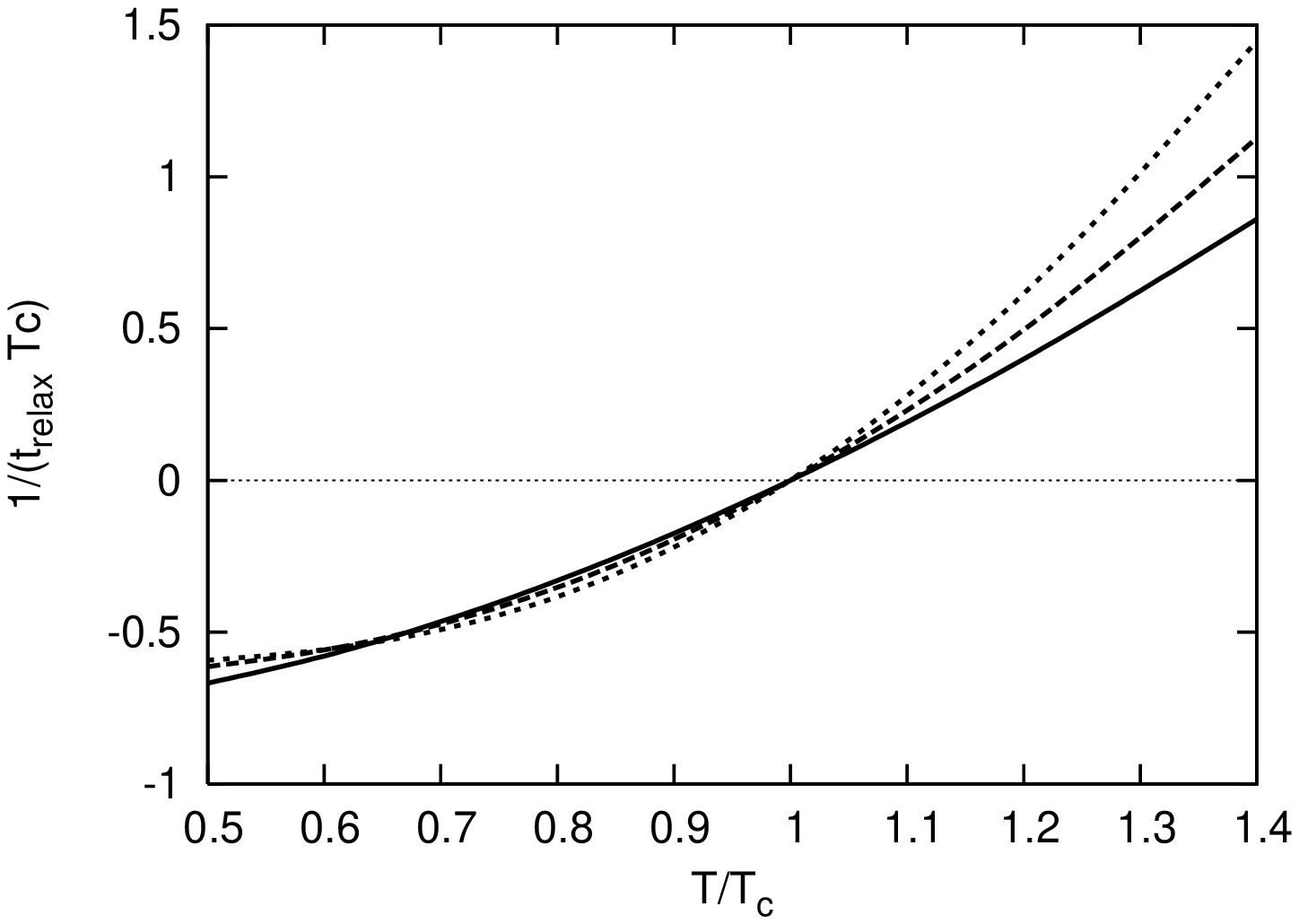}\label{fig:trela_a}}
  \subfigure[Superconducting phase]{\includegraphics[scale=0.52,angle=270]{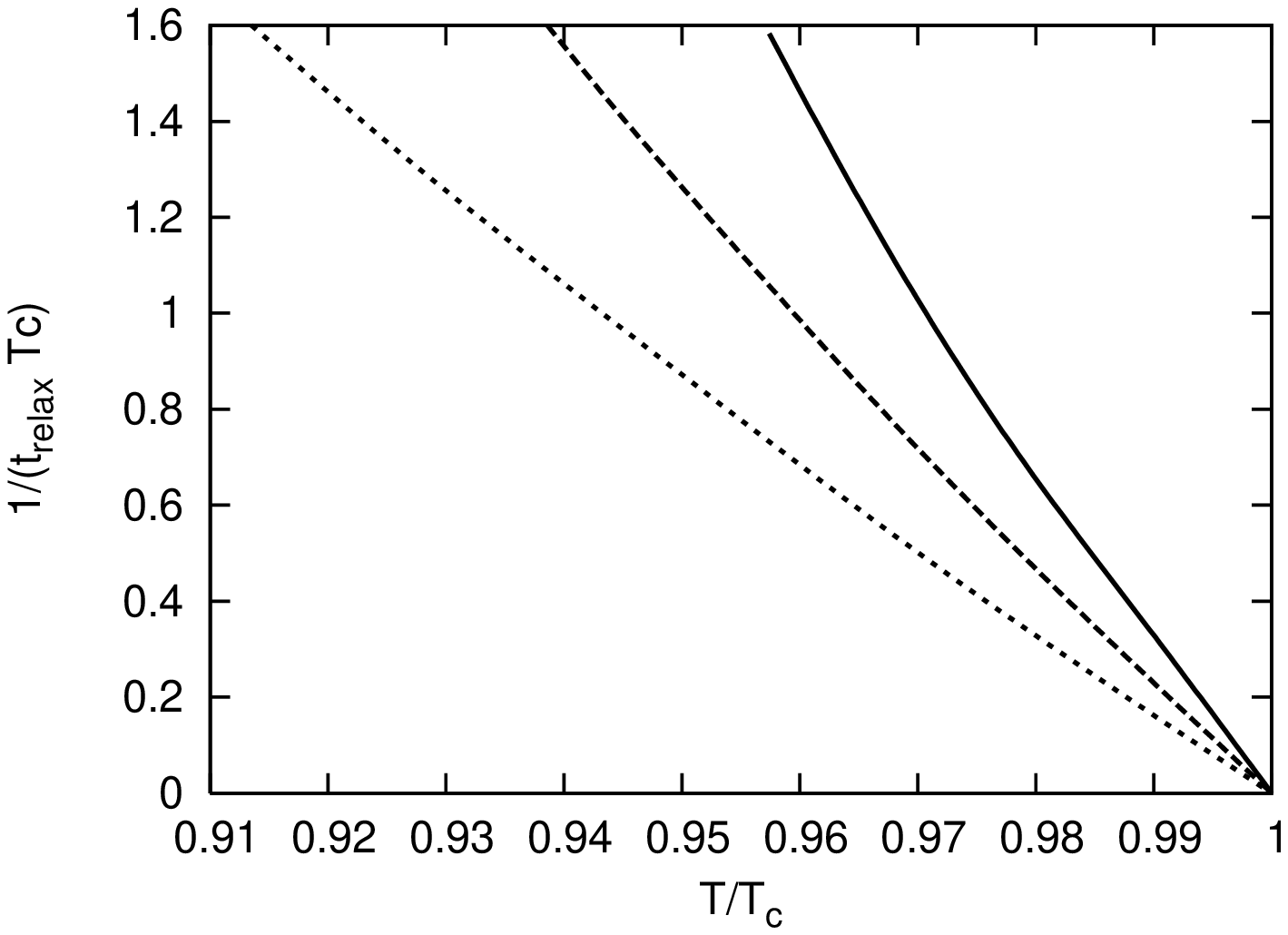}\label{fig:trela_b}}
  \caption{%
The relaxation times $t_\text{relax}$ of the order parameter against the 
temperature $T$.
Figure~(a) and (b) shows the relaxation time 
 around the normal and the condensed phases, respectively. 
The solid, dashed and doted curves correspond to $q=1.0$, $1.5$ and
   $2.0$.
In Figure~(b), we used the temperature of the final state of the time
evolution, that is, the hairy black hole.}
\end{figure}

In this subsection, we estimate the growth and decay rates of the 
order parameter.
At the beginning of the time evolution, namely, 
around the normal phase $\langle \mathcal{O}_2(t) \rangle = 0$, 
we fit the time dependence of the order parameter as
\begin{equation}
 |\langle \mathcal{O}_2(t) \rangle | =
  C\exp\left(-t/t_\text{relax}\right) \ ,
\end{equation}
where $C$ and $t_\text{relax}$ are constants and $t_\text{relax}$
represents the relaxation time scale.
In Figure~\ref{fig:trela_a}, we depict the
$1/t_\text{relax}$ around the normal phase against the temperature
for $q=1.0$, $1.5$ and $2.0$.
For $T<T_c$, $1/t_\text{relax}$ has negative value since the initial
Reissner-Nordstr\"{o}m-AdS black holes are unstable. At the onset of the
instability $T=T_c$, the $1/t_\text{relax}$ becomes zero.
It seems that, for $T/T_c \to 0$, $1/t_\text{relax}$ dose not
diverge, but converges to finite value.

For $T<T_c$, the order parameter condenses into the non-trivial value.
Around the condensed phase, we fit the time dependence of the order parameter as
\begin{equation}
 |\langle \mathcal{O}_2(t) \rangle | =
  C_1\exp\left(-t/t_\text{relax}\right) + C_2\ ,
\end{equation}
where $C_1$ and $C_2$ are constants.
In Figure~\ref{fig:trela_b}, we depict the
$1/t_\text{relax}$ around the condensed phase against the temperature.
Note that we used  in this figure the temperature of the final state of the time
evolution, that is, the hairy black hole.
We find that, at $T\to T_c$, $1/t_\text{relax}$ approaches zero.
On the other hand, for $T/T_c\to 0$, $1/t_\text{relax}$ has larger value
as the temperature becomes lower.

In the gravity side, the $1/t_\text{relax}$ is nothing but the
imaginary part of the quasinormal frequency of the fundamental mode. For
the normal 
phase, which is the Reissner-Nordstr\"{o}m-AdS black hole, quasinormal modes
are studied in~\cite{Amado:2009ts,Konoplya:2002ky,Miranda:2008vb}.
For the condensed
phase, which is the hairy black hole, quasinormal modes are studied in the
decoupling limit $q\to \infty$ in~\cite{Amado:2009ts}. 
The result in Figure~\ref{fig:trela_b}
gives the quasinormal frequency of the hairy black holes
with the back-reaction on the gravitational field.


\subsection{Evolution of horizons}
\label{Sec:horizon}

\begin{figure}[htbp]
  \begin{center}
      \includegraphics[height=9cm,angle=270,clip]{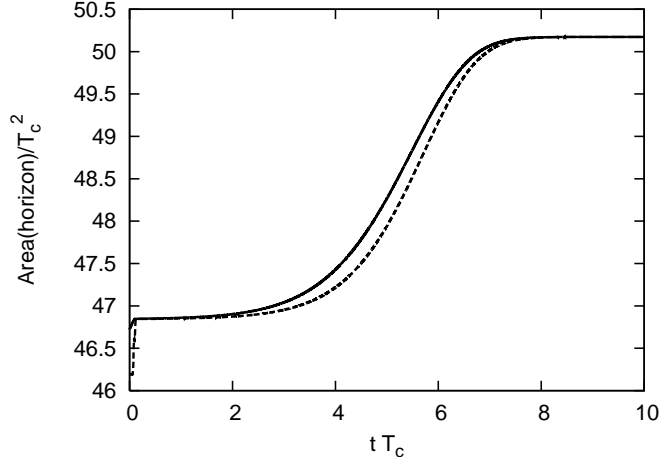}
    \caption{\label{fig:hors}%
The time evolution of the area of event and apparent horizons are
   depicted for 
$q=1.5$ and $T/T_c=0.4$ at the initial state.
Solid and dashed curves correspond to the area of event and apparent
   horizons respectively.}
  \end{center}
\end{figure}

Now, we investigate the time evolution of apparent and event horizons.
The apparent horizon $z=z_\text{AH}(t)$ can be determined from
\begin{equation}
 D\Phi(t, z_\text{AH}(t))=0\ .
\end{equation}
We determine the event horizon $z=z_\text{EH}(t)$ as follows.
For sufficiently late time, spacetimes settle static solutions.
Thus,  at late time, the event horizon can be easily determined by
$F(t,z_\text{EH}(t))=0$.
To determine the event horizon for any $t$, 
we solve the null geodesic equations,
\begin{equation}
 \dot{z}_\text{EH}(t)=-\frac{1}{2} F(t,z_\text{EH}(t))\ ,
\end{equation}
backward along the tangent to the event horizon at late time.
Then, we obtain null geodesic generators of the event horizon and find
the location of the event horizon $z=z_\text{EH}(t)$.
Once we know the $z_\text{EH}(t)$ and $z_\text{AH}(t)$, we can calculate 
the area of event and apparent horizons as
\begin{equation}
 \text{Area(event/apparent horizon)} =\Phi(t,z_\text{EH/AH}(t))^2\ .
\end{equation}
In Figure~\ref{fig:hors}, we depict the time evolution of the area of
the horizons. We find that the area of the horizons monotonically increase by the
time evolution and the event horizon has larger area than that of
the apparent horizon. 
We also see that, after the final state has settled to equilibrium,
the apparent horizon and the event horizon coincide.
In addition, 
the two horizons seem to coincide even at
the initial state.
Thus, it is quite likely that we can regard the area of the horizon as the entropy
at the initial state.
This issue will be examined further in the following subsection.

\subsection{Bulk dynamics and bulk-boundary identification}
\label{Sec:non-staticity}

Having shown properties of the horizons in the bulk, we will 
discuss how to identify the bulk with the boundary
based on observations of the bulk field dynamics.
%
When the system is stationary, we know that the area of the black hole horizon in
the bulk corresponds to the entropy of the dual boundary theory.
However, 
there is subtlety in this correspondence 
for dynamical systems (see, for
example,~\cite{Bhattacharyya:2008xc,Kinoshita:2008dq,Figueras:2009iu}
and so on).
When the system is dynamical,
we do not know the mapping between horizons and AdS boundary {\it a priori}.
In other words, 
there exists ambiguity of the time slices with which we identify the horizon and the boundary.
The null time slices we used in this paper, which are illustrated in Figure~\ref{fig:ponchi},
is nothing but one of the uncountable set of time slices.
Thus, we cannot immediately assert that Figure~\ref{fig:hors} is equivalent to
the dynamics of the entropy in the dual theory.
In this subsection, we would like to discuss this issue of the identification 
%

\begin{figure}[htbp]
  \begin{center}
   \includegraphics[width=.48\linewidth,angle=270,clip]{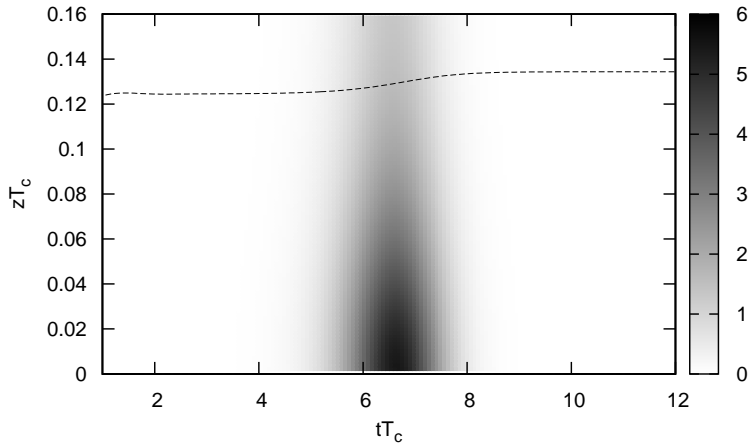}
   \caption{%
Time dependence of the bulk spacetime for the same parameters used in
   Figure~\ref{fig:hors}. The value of $\dot{\tilde\Phi}^2/(zT_c^3)^2$
   is plotted in gray-scale, in which
  the dark part represents the non-stationary region. 
In order to examine non-staticity of the entire bulk 
including the near-boundary region,
we have focused on $\dot{\tilde\Phi}/z$,
which takes finite value even on the AdS boundary,
rather than on $\dot\Phi$.
The dashed curve represents the apparent horizon.%
}
\label{fig:static}
  \end{center}
\end{figure}

Figure~\ref{fig:static} shows time dependence of a metric function
$\tilde\Phi$.
In this figure, we show the value of
$\dot{\tilde\Phi}^2/(zT_c^3)^2$ in gray-scale.
The dark region, in which $\dot{\tilde\Phi}^2/(zT_c^3)^2$ is large,
represents non-stationary region.
Rigorously speaking, this figure indicates to what extent a coordinate vector
$\xi = \partial/\partial t$ 
satisfies the Killing equation: 
$\mathcal L_\xi g_{\mu\nu} = \nabla_\mu\xi_\nu + \nabla_\nu\xi_\mu = 0$.
The quantity $|\mathcal L_\xi g_{\mu\nu}|^2 = 8{\dot\Phi}^2/\Phi^2$ is, however,
not always a good indicator of non-equilibrium property
since 
$\xi$ coincides with the timelike Killing vector on the AdS boundary and 
$|\mathcal L_\xi g_{\mu\nu}|^2$ tends to zero there.
In order to examine non-staticity even at the
boundary, we focus on the value of $\dot{\tilde\Phi}/z$ instead of
$\dot\Phi$ (see Eq.~(\ref{coeff})).
As we mentioned before, the initial state and the final state are close to
equilibrium.
It turns out that this fact appears also in Figure~\ref{fig:static}.
Another point we should note is that
the shaded region extends almost vertically in this figure.
Since we are using null time-slicing, the above fact means that the
non-stationary region extends along the ingoing null direction from the
boundary to the horizon in the bulk spacetime.
Consequently, it may be reasonable to suppose that the mapping between
quantities of the boundary and the horizon is given by 
the null time slices in the current case.
It will be interesting to examine whether we can apply this idea to other dynamical systems.

\section{Conclusions}

We studied the non-equilibrium condensation process in the holographic
superconductor by solving the time evolutions of the
Einstein-Maxwell-charged scalar system in the asymptotically AdS spacetime.
We considered small perturbations on the Reissner-Nordstr\"{o}m-AdS
black holes, which are static solutions of this system, as the initial states.
We found that, when the temperature is lower than the critical
temperature $T_c$,  initial perturbations grow exponentially and the
spacetimes settle into the hairy black holes obtained in~\cite{Hartnoll:2008kx}.
It is concluded that the hairy black holes are the final states for the
instability of the low-temperature Reissner-Nordstr\"{o}m-AdS black holes.
We also found that the hairy black holes are stable against the plane-symmetric
perturbations.
As for the superconducting order parameter in the boundary theory,
we clarified how it evolve in non-equilibrium process during the phase transition.
As a byproduct,
we obtained the relaxation time scale of the order parameter, 
which is inverse of the imaginary part of the fundamental quasi-normal mode.
Finally, we studied the time evolution of the event and apparent
horizons and discussed the their relevance to the entropy of the
boundary theory.

There are many extensions and open issues.
One of them is perturbation of dynamical spacetimes obtained in this
paper.
The dynamics of transport coefficients, such as electrical
conductivity or shear viscosity, can be known from the perturbations on the dynamical
spacetimes.
From the Dirac field perturbation, we may be able to study the
Fermi surface properties
of 
superconductors~\cite{Wu:2008zz,Chen:2009pt,Faulkner:2009am,Gubser:2009dt,Hartman:2010fk}.
These studies might deepen our understanding on dynamical phenomena in the 
AdS/CFT correspondence as well as on unknown properties of condensed matter system in 
non-equilibrium state.

In our calculations, we imposed the plane symmetry, that is, 
we assumed that the space on $(x,y)$-plane is homogeneous 
and isotropic, as in Eq.~(\ref{plane_sym}).
It is interesting to relax this symmetry.
For example, we can consider the spacetime with homogeneous and
anisotropic $(x,y)$-plane as
\begin{equation}
 ds^2=-\frac{1}{z^2}\left(F(t,z)dt^2+2dtdz\right)
+\Phi_1(t,z)^2dx^2 + \Phi_2(t,z)^2dy^2\ .
\end{equation}
The equations of motion for this metric ansatz are given by $(1+1)$-dimensional
PDEs, and thus we can straightforwardly apply
the formalism of this paper. 
In this setting,
we can take into account homogeneous electric or magnetic fields along the $x$
or $y$-directions and study the dynamics of the current
in the boundary theory.
More challenging problem is to take into
account inhomogeneity along one direction, e.g., the $y$-direction. 
The equations of motion are given by $(1+2)$-dimensional PDEs.
In the setting,
%
we may holographically realize 
a system in which 
the temperature is not homogeneous and 
the condensed phase and the non-condensed 
phase coexist.
In such an inhomogeneous system, 
we may observe the thermoelectric phenomena in superconductors~\cite{Galperin},
such as Pertier effect or 
the Zeebeck effect, or the dynamics of the surface between the two 
phases~\cite{Ozen1,Ozen2}.
Another subject which can be studied by $(1+2)$-dimensional PDEs 
is the vortex solution obtained in~\cite{Albash:2009iq,Montull:2009fe}:
we must consider the $U(1)$-symmetric spacetime where the $U(1)$ is the
rotational symmetry around the vortex.
The spacetime of the vortex solution has a $U(1)$-symmetry, which is the 
rotational symmetry around the vortex, and then 
the equations of motion are again given by $(1+2)$-dimensional PDEs.
Investigation on the non-equilibrium phenomena 
in these inhomogeneous system
and their comparison with 
theories and experiments in the field of the condensed matter physics may be interesting.

Ultimately, we should study dynamics of spacetimes without any symmetry.
Such a task may be tackled only by numerical relativity technique, which is 
developing rapidly these days~\cite{nr1,nr2,nr3,Zilhao:2010sr,nr4}.
To apply this technique to asymptotically AdS spacetime, we have to modify the formalism for 
asymptotically flat spacetime to accommodate the negative cosmological constant.
If such a technique is established, we will be able to investigate on many interesting 
subjects concerning the non-equilibrium condensation process in the AdS/CFT.
For example,
it is expected that vortex lattice~\cite{Maeda:2009vf} are formed in a dynamical system
from the simulation of the (non-holographic) superconducting system~\cite{Mondello:1990zz,Liu:1991zzc,Mondello:1992zz,Korutcheva:1998,Stephens:2001fv}.
By simulating its holographic dual in the gravity side using the technique, we 
will be able to confirm further the AdS/CFT correspondence in a dynamical setup 
and also may clarify unknown properties of dynamical strongly-coupled systems.
We believe that such studies will be fruitful for both of the quantum field 
theory and general relativity, and our study in this paper will
serve as a first step toward such ambitious future issues.

%

 \section*{Acknowledgments}
 We would like to thank Kengo~Maeda, Shinji~Mukohyama, Harvey~S.~Reall, Jiro~Soda and
 Takahiro~Tanaka for
 valuable comments on this work.
 We would also like to thank Paul~M.~Chesler 
 for kindly teaching us his numerical method
 in \cite{Chesler:2009cy}.
 KM is supported by a grant for research abroad by the JSPS (Japan).
 SK is the Yukawa Fellow and this work is partially supported by Yukawa
 Memorial Foundation.
 This work was supported by the Grant-in-Aid for the Global COE Program "The Next Generation of Physics, Spun from Universality and Emergence" from the Ministry of Education, Culture, Sports, Science and Technology (MEXT) of Japan.


\begin{thebibliography}{99}

\bibitem{Ma}
  J.~M.~Maldacena,
  ``The large N limit of superconformal field theories and supergravity,''
  Adv.\ Theor.\ Math.\ Phys.\  {\bf 2}, 231 (1998)
  [Int.\ J.\ Theor.\ Phys.\  {\bf 38}, 1113 (1999)]
  [arXiv:hep-th/9711200].

\bibitem{GKP}
  S.~S.~Gubser, I.~R.~Klebanov and A.~M.~Polyakov,
  ``Gauge theory correlators from non-critical string theory,''
  Phys.\ Lett.\  B {\bf 428}, 105 (1998)
  [arXiv:hep-th/9802109].

\bibitem{Wi}
  E.~Witten,
  ``Anti-de Sitter space and holography,''
  Adv.\ Theor.\ Math.\ Phys.\  {\bf 2}, 253 (1998)
  [arXiv:hep-th/9802150].

\bibitem{Gubser:2008px}
  S.~S.~Gubser,
  ``Breaking an Abelian gauge symmetry near a black hole horizon,''
  Phys.\ Rev.\  D {\bf 78}, 065034 (2008)
  [arXiv:0801.2977 [hep-th]].

\bibitem{Hartnoll:2008vx}
  S.~A.~Hartnoll, C.~P.~Herzog and G.~T.~Horowitz,
  ``Building a Holographic Superconductor,''
  Phys.\ Rev.\ Lett.\  {\bf 101}, 031601 (2008)
  [arXiv:0803.3295 [hep-th]].

\bibitem{Hartnoll:2008kx}
  S.~A.~Hartnoll, C.~P.~Herzog and G.~T.~Horowitz,
  ``Holographic Superconductors,''
  JHEP {\bf 0812}, 015 (2008)
  [arXiv:0810.1563 [hep-th]].

\bibitem{Mondello:1990zz}
  M.~Mondello and N.~Goldenfeld,
  ``Scaling and vortex dynamics after the quench of a system with a continuous
  symmetry,''
  Phys.\ Rev.\  A {\bf 42}, 5865 (1990).

\bibitem{Liu:1991zzc}
  F.~Liu, M.~Mondello and N.~Goldenfeld,
  ``Kinetics Of The Superconducting Transition,''
  Phys.\ Rev.\ Lett.\  {\bf 66}, 3071 (1991).

\bibitem{Mondello:1992zz}
  M.~Mondello and N.~Goldenfeld,
  ``Scaling and vortex-string dynamics in a three-dimensional system with a
  continuous symmetry,''
  Phys.\ Rev.\  A {\bf 45}, 657 (1992).

\bibitem{Korutcheva:1998}
  E.~Korutcheva and F.~Javier~de~la~Rubia,
  ``Dynamical properties of the Landau-Ginzburg
  Phys.\ Rev.\ B {\bf 58}, 5153 (1998)

\bibitem{Stephens:2001fv}
  G.~J.~Stephens, L.~M.~A.~Bettencourt and W.~H.~Zurek,
  ``Critical dynamics of gauge systems: Spontaneous vortex formation in 2D
  superconductors,''
  Phys.\ Rev.\ Lett.\  {\bf 88}, 137004 (2002)
  [arXiv:cond-mat/0108127].

\bibitem{Janik:2005zt}
  R.~A.~Janik and R.~B.~Peschanski,
  ``Asymptotic perfect fluid dynamics as a consequence of AdS/CFT,''
  Phys.\ Rev.\  D {\bf 73}, 045013 (2006)
  [arXiv:hep-th/0512162].

\bibitem{Kovtun:2004de}
  P.~Kovtun, D.~T.~Son and A.~O.~Starinets,
  ``Viscosity in strongly interacting quantum field theories from black hole
  physics,''
  Phys.\ Rev.\ Lett.\  {\bf 94}, 111601 (2005)
  [arXiv:hep-th/0405231].

\bibitem{Bhattacharyya:2008jc}
  S.~Bhattacharyya, V.~E.~Hubeny, S.~Minwalla and M.~Rangamani,
  ``Nonlinear Fluid Dynamics from Gravity,''
  JHEP {\bf 0802}, 045 (2008)
  [arXiv:0712.2456 [hep-th]].

\bibitem{Grumiller:2008va}
  D.~Grumiller and P.~Romatschke,
  ``On the collision of two shock waves in AdS5,''
  JHEP {\bf 0808}, 027 (2008)
  [arXiv:0803.3226 [hep-th]].

\bibitem{Gubser:2008pc}
  S.~S.~Gubser, S.~S.~Pufu and A.~Yarom,
  ``Entropy production in collisions of gravitational shock waves and of heavy
  ions,''
  Phys.\ Rev.\  D {\bf 78}, 066014 (2008)
  [arXiv:0805.1551 [hep-th]].

\bibitem{AlvarezGaume:2008fx}
  L.~Alvarez-Gaume, C.~Gomez, A.~Sabio Vera, A.~Tavanfar and M.~A.~Vazquez-Mozo,
  ``Critical formation of trapped surfaces in the collision of gravitational
  shock waves,''
  JHEP {\bf 0902}, 009 (2009)
  [arXiv:0811.3969 [hep-th]].

\bibitem{Lin:2009pn}
  S.~Lin and E.~Shuryak,
  ``Grazing Collisions of Gravitational Shock Waves and Entropy Production in
  Heavy Ion Collision,''
  Phys.\ Rev.\  D {\bf 79}, 124015 (2009)
  [arXiv:0902.1508 [hep-th]].

\bibitem{Gubser:2009sx}
  S.~S.~Gubser, S.~S.~Pufu and A.~Yarom,
  ``Off-center collisions in AdS$_5$ with applications to multiplicity estimates
  in heavy-ion collisions,''
  JHEP {\bf 0911}, 050 (2009)
  [arXiv:0902.4062 [hep-th]].

\bibitem{Bhattacharyya:2009uu}
  S.~Bhattacharyya and S.~Minwalla,
  ``Weak Field Black Hole Formation in Asymptotically AdS Spacetimes,''
  JHEP {\bf 0909}, 034 (2009)
  [arXiv:0904.0464 [hep-th]].

\bibitem{Chesler:2009cy}
  P.~M.~Chesler and L.~G.~Yaffe,
  ``Horizon formation and far-from-equilibrium isotropization in supersymmetric
  Yang-Mills plasma,''
  Phys.\ Rev.\ Lett.\  {\bf 102}, 211601 (2009)
  [arXiv:0812.2053 [hep-th]].

\bibitem{Amado:2009ts}
  I.~Amado, M.~Kaminski and K.~Landsteiner,
  ``Hydrodynamics of Holographic Superconductors,''
  JHEP {\bf 0905}, 021 (2009)
  [arXiv:0903.2209 [hep-th]].

\bibitem{Hartnoll:2009sz}
  S.~A.~Hartnoll,
  ``Lectures on holographic methods for condensed matter physics,''
  Class.\ Quant.\ Grav.\  {\bf 26}, 224002 (2009)
  [arXiv:0903.3246 [hep-th]].

\bibitem{Horowitz:2010gk}
  G.~T.~Horowitz,
  ``Introduction to Holographic Superconductors,''
  arXiv:1002.1722 [hep-th].

\bibitem{Maeda:2009wv}
  K.~Maeda, M.~Natsuume and T.~Okamura,
  ``Universality class of holographic superconductors,''
  Phys.\ Rev.\  D {\bf 79}, 126004 (2009)
  [arXiv:0904.1914 [hep-th]].

\bibitem{Herzog:2008he}
  C.~P.~Herzog, P.~K.~Kovtun and D.~T.~Son,
  ``Holographic model of superfluidity,''
  Phys.\ Rev.\  D {\bf 79}, 066002 (2009)
  [arXiv:0809.4870 [hep-th]].

\bibitem{Herzog:2009md}
  C.~P.~Herzog and A.~Yarom,
  ``Sound modes in holographic superfluids,''
  Phys.\ Rev.\  D {\bf 80}, 106002 (2009)
  [arXiv:0906.4810 [hep-th]].

\bibitem{Maeda:2010hf}
  K.~Maeda, J.~i.~Koga and S.~Fujii,
  ``The final fate of instability of Reissner-Nordstr\'om-anti-de Sitter black
  holes by charged complex scalar fields,''
  arXiv:1003.2689 [gr-qc].

\bibitem{Konoplya:2002ky}
  R.~A.~Konoplya,
  ``Decay of charged scalar field around a black hole: Quasinormal modes of
  R-N, R-N-AdS and dilaton black hole,''
  Phys.\ Rev.\  D {\bf 66}, 084007 (2002)
  [arXiv:gr-qc/0207028].

\bibitem{Miranda:2008vb}
  A.~S.~Miranda, J.~Morgan and V.~T.~Zanchin,
  ``Quasinormal modes of plane-symmetric black holes according to the AdS/CFT
  correspondence,''
  JHEP {\bf 0811}, 030 (2008)
  [arXiv:0809.0297 [hep-th]].


\bibitem{Bhattacharyya:2008xc}
  S.~Bhattacharyya {\it et al.},
  ``Local Fluid Dynamical Entropy from Gravity,''
  JHEP {\bf 0806}, 055 (2008)
  [arXiv:0803.2526 [hep-th]].

\bibitem{Kinoshita:2008dq}
  S.~Kinoshita, S.~Mukohyama, S.~Nakamura and K.~y.~Oda,
  ``A Holographic Dual of Bjorken Flow,''
  Prog.\ Theor.\ Phys.\  {\bf 121}, 121 (2009)
  [arXiv:0807.3797 [hep-th]].

\bibitem{Figueras:2009iu}
  P.~Figueras, V.~E.~Hubeny, M.~Rangamani and S.~F.~Ross,
  ``Dynamical black holes and expanding plasmas,''
  JHEP {\bf 0904}, 137 (2009)
  [arXiv:0902.4696 [hep-th]].


\bibitem{Wu:2008zz}
  C.~Wu, K.~Sun, E.~Fradkin and S.~C.~Zhang,
  ``Fermi liquid instabilities in the spin channel,''
  Phys.\ Rev.\  B {\bf 75} (2007) 115103.

\bibitem{Chen:2009pt}
  J.~W.~Chen, Y.~J.~Kao and W.~Y.~Wen,
  ``Peak-Dip-Hump from Holographic Superconductivity,''
  arXiv:0911.2821 [hep-th].

\bibitem{Faulkner:2009am}
  T.~Faulkner, G.~T.~Horowitz, J.~McGreevy, M.~M.~Roberts and D.~Vegh,
  ``Photoemission 'experiments' on holographic superconductors,''
  JHEP {\bf 1003}, 121 (2010)
  [arXiv:0911.3402 [hep-th]].

\bibitem{Gubser:2009dt}
  S.~S.~Gubser, F.~D.~Rocha and P.~Talavera,
  ``Normalizable fermion modes in a holographic superconductor,''
  arXiv:0911.3632 [hep-th].

\bibitem{Hartman:2010fk}
  T.~Hartman and S.~A.~Hartnoll,
  ``Cooper pairing near charged black holes,''
  arXiv:1003.1918 [hep-th].

\bibitem{Galperin}
Y.~M. Galperin, V.~L. Gurevich, V.~I. Kozub and A.~L. Shelankov,
``Theory of thermoelectric phenomena in superconductors,''
Phys.\ Rev.\ B {\bf 65}, 064531 (2002)

\bibitem{Ozen1}
O.~Ozen and R.~Narayanan,
``The physics of evaporative and convective instabilities in bilayer systems: Linear theory,''
Physics of Fluids {\bf 16}, 4644 (2004)

\bibitem{Ozen2}
O.~Ozen and R.~Narayanan
``The physics of evaporative instability in bilayer systems: Weak nonlinear theory,''
Physics of Fluids {\bf 16}, 4653 (2004)

\bibitem{Albash:2009iq}
  T.~Albash and C.~V.~Johnson,
  ``Vortex and Droplet Engineering in Holographic Superconductors,''
  Phys.\ Rev.\  D {\bf 80}, 126009 (2009)
  [arXiv:0906.1795 [hep-th]].

\bibitem{Montull:2009fe}
  M.~Montull, A.~Pomarol and P.~J.~Silva,
  ``The Holographic Superconductor Vortex,''
  Phys.\ Rev.\ Lett.\  {\bf 103}, 091601 (2009)
  [arXiv:0906.2396 [hep-th]].

\bibitem{nr1}
  H.~Yoshino and M.~Shibata,
  ``Higher-dimensional numerical relativity: Formulation and code tests,''
  Phys.\ Rev.\  D {\bf 80} (2009) 084025
  [arXiv:0907.2760 [gr-qc]].

\bibitem{nr2}
  K.~i.~Nakao, H.~Abe, H.~Yoshino and M.~Shibata,
  ``Maximal slicing of D-dimensional spherically-symmetric vacuum spacetime,''
  Phys.\ Rev.\  D {\bf 80} (2009) 084028
  [arXiv:0908.0799 [gr-qc]].

\bibitem{nr3}
  M.~Shibata and H.~Yoshino,
  ``Nonaxisymmetric instability of rapidly rotating black hole in five
  dimensions,''
  Phys.\ Rev.\  D {\bf 81} (2010) 021501
  [arXiv:0912.3606 [gr-qc]].

\bibitem{Zilhao:2010sr}
  M.~Zilhao, H.~Witek, U.~Sperhake, V.~Cardoso, L.~Gualtieri, C.~Herdeiro and A.~Nerozzi,
  ``Numerical relativity for D dimensional axially symmetric space-times:
  formalism and code tests,''
  Phys.\ Rev.\  D {\bf 81}, 084052 (2010)
  [arXiv:1001.2302 [gr-qc]].

\bibitem{nr4}
  H.~Witek, V.~Cardoso, C.~Herdeiro, A.~Nerozzi, U.~Sperhake and M.~Zilhao,
  ``Black holes in a box: towards the numerical evolution of black holes in
  AdS,''
  arXiv:1004.4633 [hep-th].

\bibitem{Maeda:2009vf}
  K.~Maeda, M.~Natsuume and T.~Okamura,
  ``Vortex lattice for a holographic superconductor,''
  Phys.\ Rev.\  D {\bf 81}, 026002 (2010)
  [arXiv:0910.4475 [hep-th]].


\end{thebibliography}
\end{document}